\documentclass[a4paper]{jpconf}
\usepackage{graphicx}

\usepackage{graphics,graphicx}
\usepackage{multicol} 
\usepackage{parskip}
\usepackage{amsmath}
\usepackage{multirow}

\begin{document}
\title{Impact of xenon doping in the scintillation light in a large liquid-argon TPC}

\author{J. Soto-Oton on behalf of the DUNE Collaboration}

\address{CIEMAT, Av. Complutense 40, 28040 Madrid, Spain}

\ead{j.soto@cern.ch}

\begin{abstract}
The use of xenon-doped liquid argon (Xe-doped LAr) is a promising alternative for large-scale liquid argon Time Projection Chambers (LAr-TPC), since it mitigates the light suppression due to impurities and it also improves the photon-detection efficiency and uniformity with the distance. This study analyses the impact of using Xe-doped LAr in ProtoDUNE Dual-Phase, a 750 ton Dual-Phase LAr-TPC placed at CERN. ProtoDUNE Dual-Phase completed a Xe-doping data-taking campaign in summer 2020 by re-filling the detector with 230 tons of Xe-doped LAr contaminated with nitrogen, and performing dedicated nitrogen injections. The effects of the presence of Xe at 5.8\,ppm in the scintillation light production and propagation are analysed in this paper, showing an increase of the collected photons, but a suppression of the light signal amplitude. A 60\% increase of the light attenuation length is measured. The impact on the scintillation time profile is also studied. A model to fit the time profile is proposed and the time constants of the physics processes are obtained.
\end{abstract}

\section{Introduction}

In the presence of xenon, liquid-argon (LAr) scintillation light is shifted from 127\,nm to longer wavelengths, reducing the quenching introduced by impurities. Additionally, the longer Rayleigh Scattering Length (RSL) for the new photons enhances the light detection with the distance and improves the detection uniformity. ProtoDUNE Dual-Phase (DP) is a 750 t LAr time projection chamber (LAr-TPC) placed at the CERN Neutrino Platform, and it took data from June 2019 to September 2020 using pure LAr and xenon-doped LAr. ProtoDUNE-DP uses 36 Hamamatsu R5912-20Mod PMTs placed at the bottom to detect scintillation light from cosmic muons \cite{protoDUNEPMTs}. ProtoDUNE-DP is a prototype of the Deep Underground Neutrino Experiment (DUNE) to validate the technology.

DUNE \cite{DUNEtdrv1} will be a dual-site experiment that aims at measuring the neutrino-oscillation parameters with a precision that will allow to determine the CP violation phase and the neutrino mass ordering \cite{DUNEOscillations}. It will also perform nucleon decay searches \cite{DUNEBSM} and it will detect astrophysical neutrinos from a core-collapse supernova within the galaxy \cite{DUNESN}. DUNE will consist of a neutrino beam and a near detector placed at Fermilab, and a far detector situated underground at the Sanford Underground Research Facility (SURF) 1300 km away from Fermilab.

In July 2020, after the operation with pure LAr, ProtoDUNE Dual-Phase was re-filled with 230 tons of Xe-doped LAr contaminated with nitrogen from ProtoDUNE Single-Phase \cite{Abi_2020}. After the filling, dedicated N$_{2}$ injections were performed in order to study the light attenuation. Data were taken during all the process. As a result, four situations with different concentrations of Xe and N$_{2}$ with the same volume of liquid are under study. They are summarized in Tab.~\ref{tab:XenonAttenuation}.

\begin{figure}[htp]
    \centering
    \includegraphics[width=0.65\textwidth]{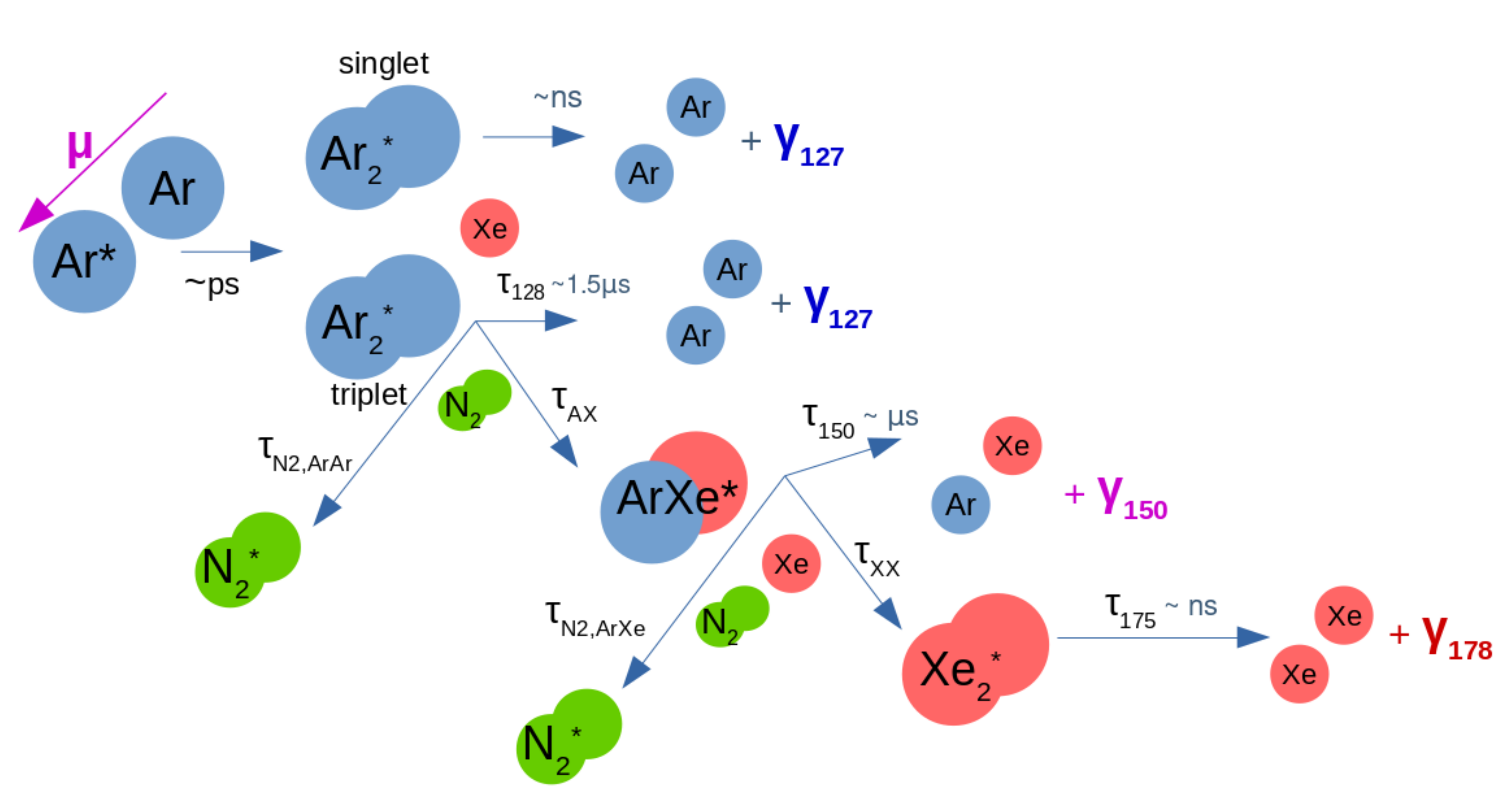}
    \caption{Diagram of the processes involved in the scintillation light production of Ar, Xe and N$_{2}$ mixtures.}
    \label{fig:ScintillationDiagram}
    \end{figure}

The physical processes involved in the scintillation light production in Xe-doped LAr contaminated with N$_{2}$ are collected in Tab.~\ref{tab:ScintModelResults1} and Fig.~\ref{fig:ScintillationDiagram}. First, the excitation produced by a crossing muon produces $Ar_{2}^{\ast}$ in two levels of energy, singlet and triplet. While the excimers in the single state decay fastly producing 127-nm photons, the triple-state ones can either decay (producing 127-nm photons), be quenched by the nitrogen or produce $ArXe^{\ast}$. This $ArXe^{\ast}$ can either decay (producing 150-nm photons), be quenched again by nitrogen or produce $Xe_{2}^{\ast}$. Finally, $Xe_{2}^{\ast}$ excimers decay very fast producing 178-nm photons.

Being the scintillation light in pure LAr composed only by 127\,nm photons, in the presence of xenon the scintillation light is not monochromatic anymore, having three wavelengths: The fast component will remain at 127\,nm, and the slow will be composed by 127, 150 and 178\,nm photons.

\section{Results}

The variation of the average scintillation light detected (called S1 signal) is studied under the four doping situations with two types of events: PMT-trigger events, which correspond to muons leaving a signal of amplitude larger than 27 photoelectrons on a PMT placed in the center of the detector, and CRT-trigger events, which are muons crossing diagonally the detector, triggered by the Cosmic Ray Taggers (CRTs), placed at the cryostat walls. PMT-trigger muon tracks are at an average distance of $\sim$2 m from the trigger PMT, while the PMT-track distance for CRT trigger muons is in the range of 3-5 m. The 11 PMTs placed close to the trigger PMT have been selected for the PMT trigger data, to focus on PMTs at a similar distance to the track. The 18 PMTs placed below the track selected by the CRTs are considered, since the level of detected light in the other PMTs is reduced, and dominated by the background light. The selection is draw in the small diagram in Fig.~\ref{fig:LightYieldFilled}.

\subsection{Impact on the average S1 charge and S1 amplitude}
\begin{figure}[h]
\begin{minipage}{18pc}
\includegraphics[width=15pc]{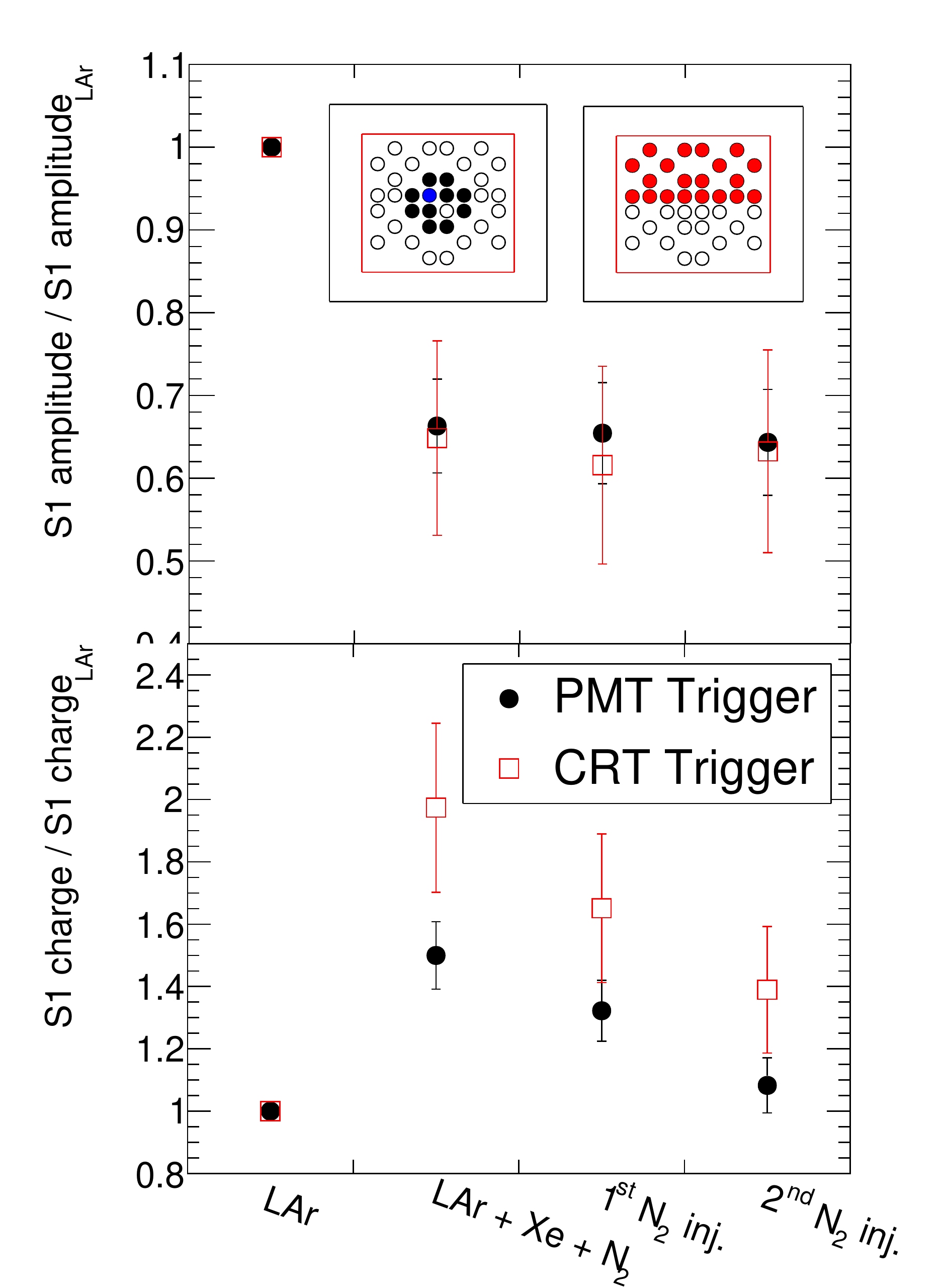}
\caption{\label{fig:LightYieldFilled}Average S1 amplitude (top) and S1 charge (bottom) in the four doping situations relative to pure LAr. PMT (CRT) trigger events in black (red). PMT selection is shown in the diagrams with its geometrical position in the detector.}
\end{minipage}\hspace{2pc}%
\begin{minipage}{18pc}
\includegraphics[width=17pc]{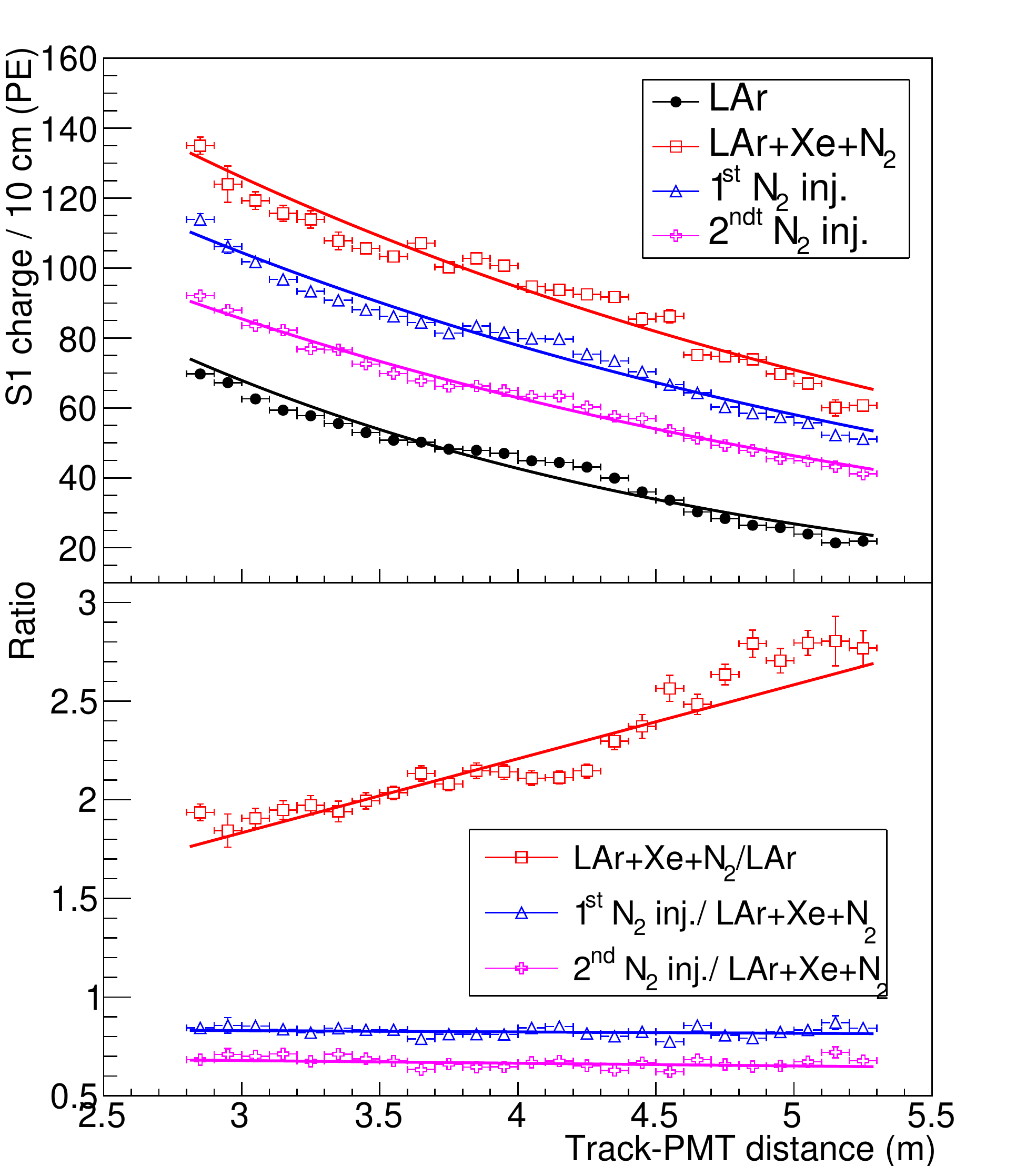}
\caption{\label{fig:CRT_ChargeVsDistance}Top panel: S1 charge vs PMT-Track distance for the 4 doping situations. Each data-set is fitted to an exponential, and results are shown in Tab.~\ref{tab:XenonAttenuation}. Bottom panel: Ratio between the data shown in the top panel.}
\end{minipage} 
\end{figure}

Figure ~\ref{fig:LightYieldFilled} shows the average variation of the S1 amplitude and S1 charge with respect to the pure LAr. A similar decrease in the fast component (S1 amplitude) of 35\% is observed in both triggers when adding Xe. This decrease is due to the photo-absorption at 127\,nm as reported in \cite{Neumeier2}. The fast component (S1 amplitude) is not affected by the N$_{2}$ injections in both triggers. This means that the photo-absorption of the 127-nm photons does not increase after the injections in opposition to the expected suppression reported by the literature \cite{Jones_N2}. Both triggers show a large increase in the S1 charge when adding xenon. The larger increase in the CRT trigger (a factor of 2 vs a factor of 1.5) compared to PMT trigger points to a better collection of light from far distances thanks to a larger RSL, as expected. Finally N$_{2}$ injections (+1\,ppm and +2\,ppm) reduce the S1 charge by $\sim$30\% in both triggers due to the N$_{2}$ quenching.

\subsection{Impact on the attenuation length}
The CRT panels allow to reconstruct the muon-track position. In this way, the dependence of the light yield with the PMT-track distance can be studied for the CRT-trigger events.

Figure \ref{fig:CRT_ChargeVsDistance} shows the dependence of the S1 charge  with the PMT-track distance in the four doping situations under study. The curves are fitted to an exponential function to estimate the attenuation length, which values are summarized in Tab.~\ref{tab:XenonAttenuation}. Results show an increase of 60\% of the attenuation length when adding Xe, as expected, while it only decreases 6\% when adding $N_{2}$.  The lower panel in Fig.~\ref{fig:CRT_ChargeVsDistance} shows the ratio of the curves. The red curve (squares) shows the improved detection uniformity when adding Xe, with a gain of a factor 3 for tracks at 5\,m vs. only a factor 2 at 3\,m. Again, this is due to the larger RSL. The flat blue (triangles) and magenta (diamonds) lines for the N$_{2}$ injections show that there is not a substantial dependence on the distance, as seen also in Fig.~\ref{fig:LightYieldFilled}, although a 6\% reduction on the attenuation length is measured in the exponential fit. This indicates a small photo-absorption due to N$_{2}$.

\begin{table}[h]
\begin{center}
\lineup
\begin{tabular}{*{4}{l}}
\br
    Situation & [Xe](ppm) & [N$_{2}$] (ppm)&  $\lambda_{att}$ (cm) \\
\mr
LAr                   & 0   & 0   &  $216\pm2$\\
LAr + Xe + N$_{2}$    & 5.8 & 1.7 & $348\pm6$\\
1$^{st}$ N$_{2}$ inj. & 5.8 & 2.7 & $341\pm4$\\
2$^{nd}$ N$_{2}$ inj. & 5.8 & 4.7 & $326\pm3$\\
\br
\end{tabular}
\caption{Attenuation lengths in the four doping situations extracted from the fits in Fig.~\ref{fig:CRT_ChargeVsDistance}.}
\label{tab:XenonAttenuation}
\end{center}
\end{table}


\subsection{Impact on the scintillation time profile}
\begin{figure}[ht]
\begin{minipage}{18pc}
\includegraphics[width=17pc]{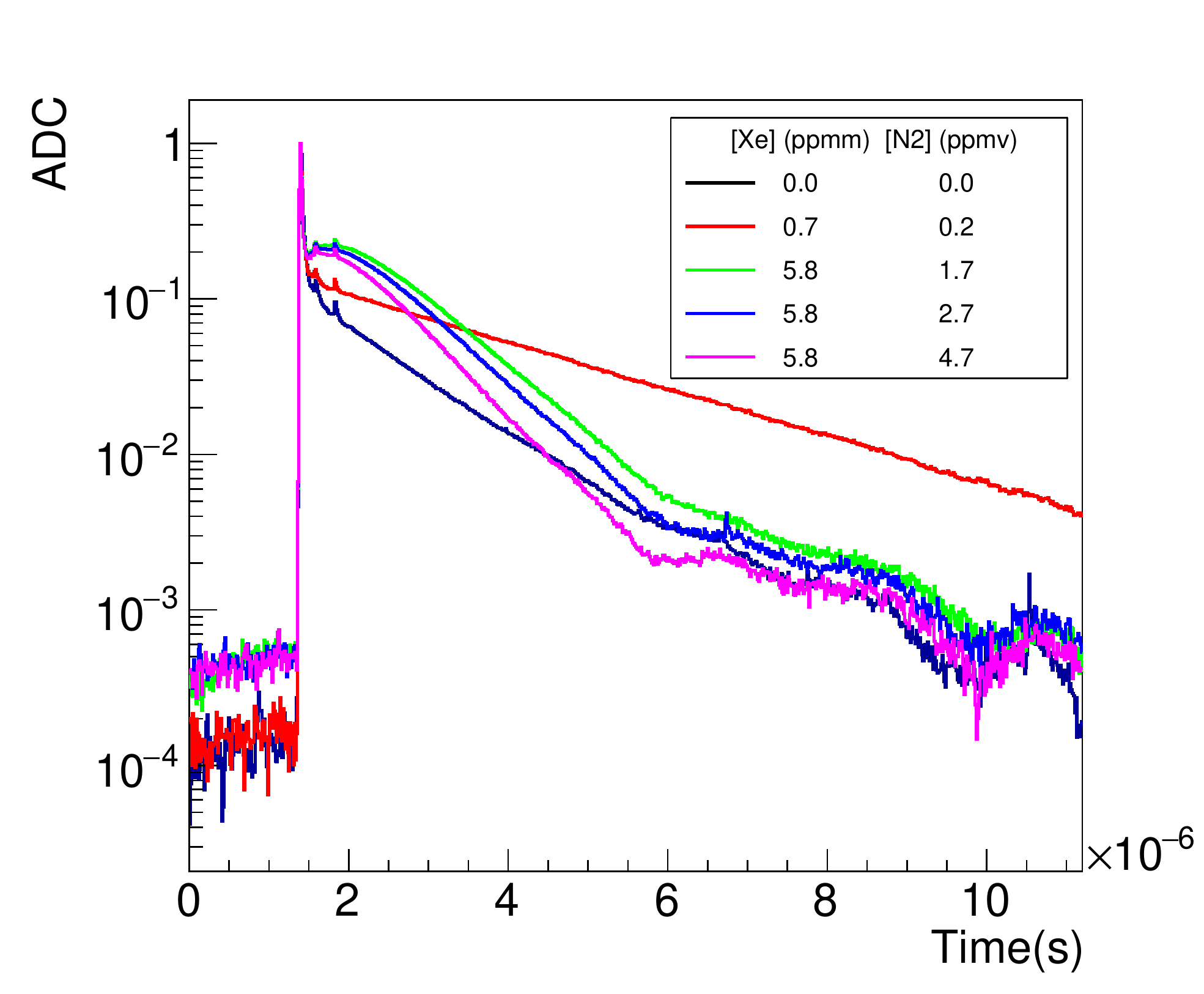}
\caption{\label{fig:TimeProfileVariation}Time profile evolution at different concentrations. Pure LAr in black, at the very beginning of the filling in red, once the detector is filled in green, blue and magenta after the N$_{2}$ injections. The sampling is 16\,ns.}
\end{minipage}\hspace{1pc}%
\begin{minipage}{18pc}
\includegraphics[width=17pc]{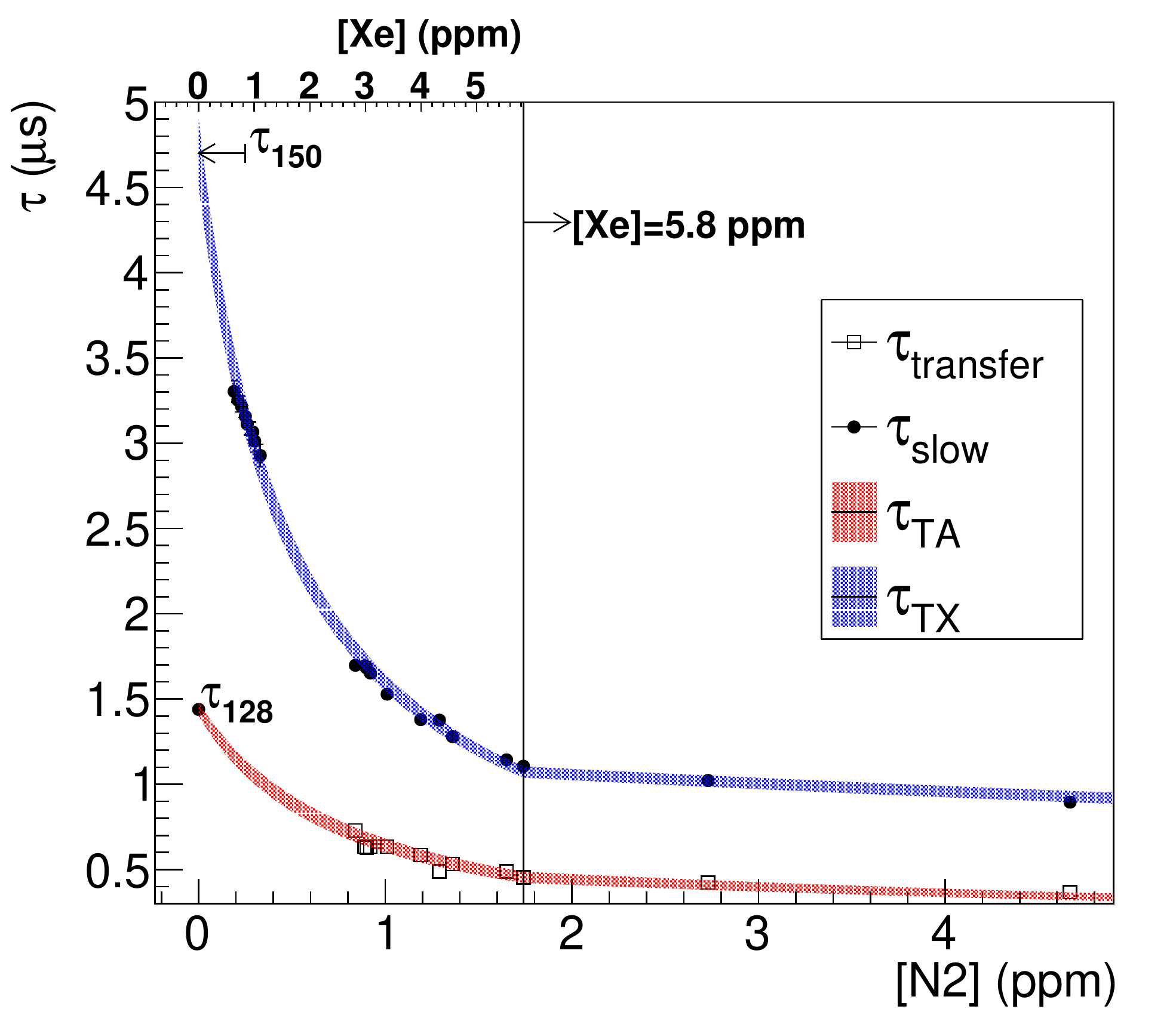}
\caption{\label{fig:GlobalFit} Time constants vs Xe (top axis) and N$_{2}$ (bottom axis) concentration. Black points show the result from the individual fits. The blue and red shaded areas show the result from the global fit $\pm 1\sigma$ to Eq.\ref{eq:ModelEq}.}
\end{minipage} 
\end{figure}

The scintillation time profile is obtained by adding waveforms from many events. Figure \ref{fig:TimeProfileVariation} shows the time profile evolution during filling and N$_{2}$ injections. The typical average waveform in pure LAr is shown in black.  First, the slow component increases largely at the very beginning of the filling as the Xe concentration is still very small (red curve).
This is due to the long lifetime of the ArXe excimer.
A second and wider bump appears as more Xe atoms are available to create excimers (green curve). Finally, the slow component shrinks back as the N$_{2}$ is injected (blue and magenta curves).

The average waveforms in Fig.~\ref{fig:TimeProfileVariation} are fitted to the sum of three exponential functions convoluted with a Gaussian to account for the PMT response. The first exponential describes the fast signal, and the second and third exponential functions model the raise ($\tau_{transfer}$) and decay ($\tau_{slow}$) of the second bump from the Xe light. The values obtained are shown in Fig.~\ref{fig:GlobalFit}.

Assuming a fast decay of the $Xe_{2}^{*}$ excimers, it can be shown that the raise of the bump ($\tau_{transfer}$) corresponds to the rate of $Ar_{2}^{*}$ disappearance ($\tau_{TA}$) and the decay of the bump to the rate of $ArXe^{*}$ disappearance ($\tau_{TX}$), depending on the Xe and N$_{2}$ concentration as follows:
    \begin{equation}
    \begin{split}
\frac{1}{\tau_{TA}} & = \frac{1}{\tau_{N2,ArAr}} +\frac{1}{\tau_{AX}} + \frac{1}{\tau_{128}}= k_{N2,ArAr} [N2] + k_{AX} [Xe] + \frac{1}{\tau_{128}} \\
\frac{1}{\tau_{TX}} & = \frac{1}{\tau_{N2,ArXe}} + \frac{1}{\tau_{XX}} + \frac{1}{\tau_{150}} = k_{N2,ArXe} [N2] + k_{XX} [Xe] + \frac{1}{\tau_{150}}
    \label{eq:ModelEq}
    \end{split}
    \end{equation}
Where $k_{i}$ are the quenching constants independent from the Xe or N$_{2}$ concentrations.

A global fit to all the time profiles is performed, extracting the physical constants from the model. Some of them are measured in this work for the first time. Fig.~\ref{fig:GlobalFit} shows the result of the fit (shaded areas) plotted together with the individual fits (points). The parameters obtained are shown in Tab.~\ref{tab:ScintModelResults1}, comparing with the reference value in the literature when possible. The time decay of the ArXe$^{*}$ excimer is measured $\tau_{150}=4.7 \pm 0.1 \mu s$. A value not far from the literature is obtained for $\tau_{XX}$, even though the Xe concentrations under study are very different.

\begin{table}[!ht]
\begin{center}
    \begin{tabular}{*{4}{c}}
    \br
    Process & Time & Results & Literature\\
    \mr
$Ar_{2}^{\ast}( {}^{3}{\Sigma}_{u}^{+} ) \rightarrow 2 Ar + \gamma \text{ (128 nm)}$ & $\tau_{128}$ ($\mu s$) & 1.44 & 
$\sim 1.6$  \cite{Hitachi} 
\\
$Ar_{2}^{\ast}( {}^{3}{\Sigma}_{u}^{+} ) + Xe \rightarrow ArXe^{\ast}+ Ar$ & $\tau_{AX}$ [Xe] ($\mu s$  ppm) & $5.4\pm0.3$ \\
$ArXe^{\ast} \rightarrow Ar + Xe + \gamma \text{ (150nm)} $ & $\tau_{150}$ ($\mu s$) & $4.7\pm0.1$ \\
$ArXe^{\ast} + Xe  \rightarrow Xe_{2}^{\ast}  ({}^{1,3}{\Sigma}_{u}^{+}) + Ar$ & $\tau_{XX}$ [Xe] ($\mu s$  ppm) & $9.2\pm0.1$ & 11.4 \cite{Wahl_2014}\\
$Ar_{2}^{\ast}( {}^{3}{\Sigma}_{u}^{+} ) + N_{2} \rightarrow 2 Ar + N_{2}$ & $\tau_{N2,ArAr}$ [N$_{2}$] ($\mu s$  ppm) & $4.1 \pm 0.1$ & $9.1\pm0.1$ \cite{Acciarri_N2} \\
$ArXe^{\ast} + N_{2}  \rightarrow Ar + Xe + N_{2}$ & $\tau_{N2,ArXe}$ [N$_{2}$] ($\mu s$  ppm)  & $20.3\pm0.7$ & \\
    \br
    \end{tabular}
    \caption{Physical time constants obtained from the global fit.}
    \label{tab:ScintModelResults1}
\end{center}
\end{table}

\section{Conclusions}
ProtoDUNE-DP data shows that Xe doping is a promising technique for large LAr-TPCs. Since it increases the collected light at large distances, it is of special interest for large-scale detectors like DUNE. With a small doping level of 5.8 ppm of Xe (and even with the presence of 1.7\,ppm of N$_{2}$), an increase of 100\% of the collected light is reported for muons crossing at a distance of 3-5\,m from the PMTs, improving the scintillation light detection efficiency. Also an increase in the attenuation length of 60\% is measured, improving the detection uniformity. However, the 35\% reduction observed in the fast signal might compromise the performance of a light-based trigger of a future xenon-doped based LAr-TPC.

\section*{Acknowledgments}
This  project  has  received  funding  from  the  European  Union  Horizon  2020  Research  and Innovation  programme  under  Grant  Agreement  no.  654168; from  the  Spanish  Ministerio  deEconomia  y  Competitividad  (SEIDI-MINECO)  under  Grant  no.  FPA2016-77347-C2-1-P  and MdM-2015-0509;  from  the  Comunidad  de  Madrid;  and  the  support  of  a  fellowship  from  ”laCaixa” Foundation (ID 100010434) with code LCF/BQ/DI18/11660043.

\section*{References}
\bibliographystyle{iopart-num}
\bibliography{biblio}

\end{document}